\documentclass{AUJarticle}
\usepackage[cmex10]{amsmath}
\usepackage[utf8x]{inputenc}
\usepackage[nocompress]{cite}
\usepackage{graphicx, multirow, booktabs, color, listings}
\usepackage{balance}

\usepackage{verbatim}
\usepackage{orcidlink}
\usepackage{hyperref}

\graphicspath{{figures/}}

\begin{document}
	\title{Safety of the Intended Functionality Concept Integration into a Validation Tool Suite}

	\addauthor{V\'ictor J. Exp\'osito Jim\'enez, Bernhard Winkler, Joaquim M. Castella Triginer} {Virtual Vehicle Research GmbH, Inffeldgasse 21A, 8010, Graz, Austria; Tel: +43 316 873 9001} {\{victor.expositojimenez,bernhard.winkler,joaquim.castellatrigine\}@v2c2.at}
	
	\addauthor{Heiko Scharke, Hannes Schneider} {AVL List GmbH, Hans-List-Platz 1, 8020, Graz, Austria; Tel: +43 316 787 0} {\{heiko.scharke, hannes.schneider\}@avl.com}
	
	\addauthor{Eugen Brenner, Georg Macher} {Graz University of Technology, Rechbauerstraße 12, 8010, Graz, Austria; Tel: +43 316 873 6401} {\{brenner, georg.macher\}@tugraz.at}

	\issuev{XX}
	\issuen{X}
	\issued{September 2023}
	
	\shortauthor{V. J. Exp\'osito Jim\'enez et al.}
	\shorttitle{SOTIF concept through a Validation Tool Suite}
	
	\thispagestyle{plain}
		
	\maketitle

\begin{abstract}	
	Nowadays, the increasing complexity of Advanced Driver Assistance Systems (ADAS) and Automated Driving (AD) means that the industry must move towards a scenario-based approach to validation rather than relying on established technology-based methods. This new focus also requires the validation process to take into account Safety of the Intended Functionality (SOTIF), as many scenarios may trigger hazardous vehicle behaviour. Thus, this work demonstrates how the integration of the SOTIF process within an existing validation tool suite can be achieved. The necessary adaptations are explained with accompanying examples to aid comprehension of the approach. 	   	  
	

\end{abstract}

\section{Introduction}

\begin{figure}
		\centering
		\includegraphics[width=0.47\textwidth]{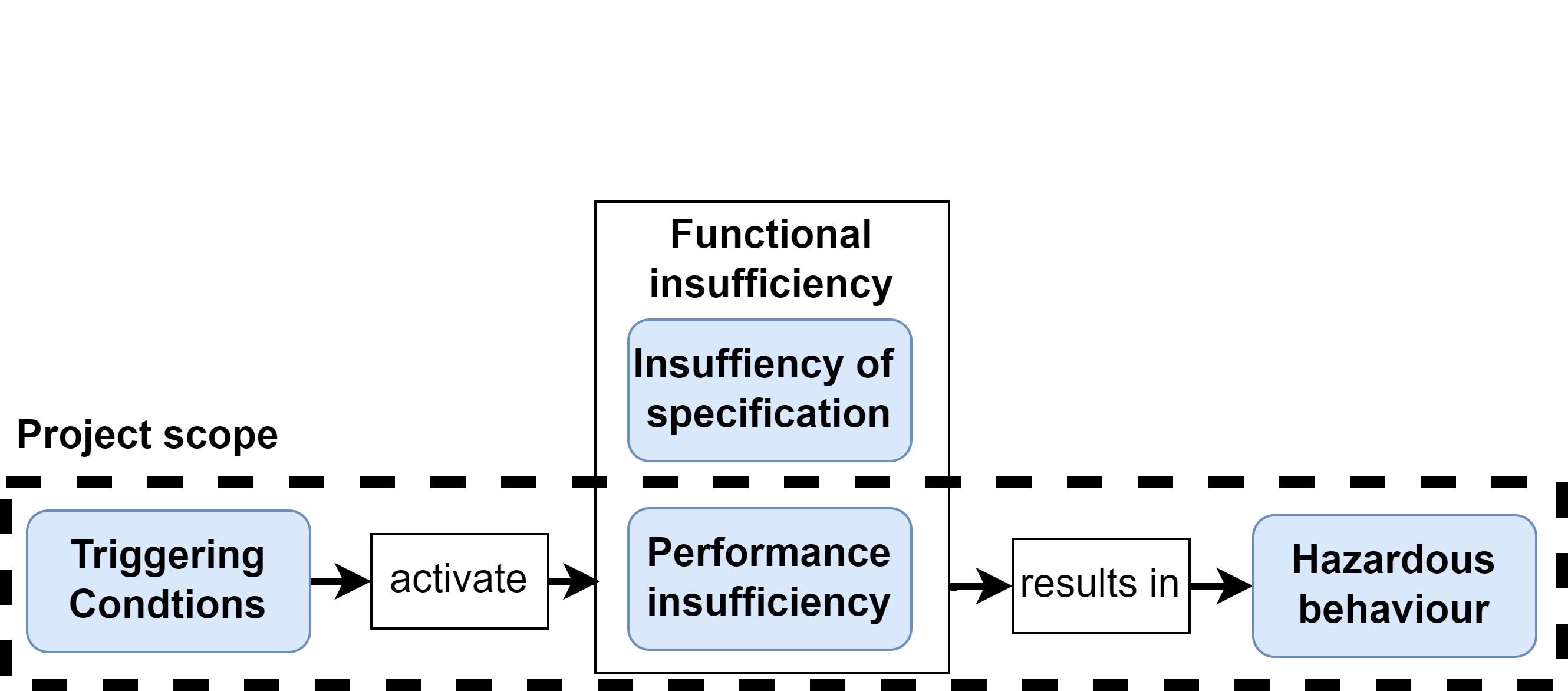}
		\caption{Cause-and-effect model from ISO21448}
		\label{fig:masa_cause_and_effect_model}
	\end{figure}

	 Scenario validation plays a significant role in the entire vehicle validation process as an increasing number of features and safety systems rely on sensors. Unlike functional safety~\cite{ISO26262} or cybersecurity~\cite{SAEJ3016}, which covers failures and malfunctions, and external attacks respectively, the Safety of the Intended Functionality (SOTIF) standard~\cite{ISO21448} focuses on the technical shortcomings and human misuses that may result in hazardous behaviour at vehicle level. Its focus is to increase the identification of hazardous scenarios to be validated as well as to minimise the area in which unknown hazardous scenarios could appear. Figure~\ref{fig:masa_cause_and_effect_model} shows the cause-and-effect model in which is depicted how a potential triggering condition could result in a hazardous behaviour at the end of the process. According to the ISO21448, a triggering condition is a \textit{“specific condition of a scenario that serves as an initiator for a subsequent system reaction contributing to either a hazardous behaviour or an inability to prevent or detect and mitigate a reasonably foreseeable indirect misuse. The concept of 'triggering' includes the possibility that there can be multiple conditions that can gradually happen, leading to hazardous behaviour or the inability to prevent or detect and mitigate a reasonably foreseeable misuse. The term “potential triggering condition” can be used when the ability to initiate a corresponding reaction is not yet established”}. Another concise definition is given in\cite{triggering_conditions}, where a triggering condition is defined as \textit{“an external condition (relative to ego-vehicle) in a scenario that triggers one or multiple functional insufficiencies and further results in hazardous behaviour. They are system-dependent was well”}. The SOTIF standard also defines a performance insufficiency as a \textit{“limitation of the technical capability contributing to a hazardous behaviour or inability to prevent or detect and mitigate reasonably foreseeable indirect misuse when activated by one or more triggering conditions}. Examples of performance insufficiencies could be the limitation of the actuation or the perception range of the sensor used detect objects. Consequently, a functional insufficiency is defined as an \textit{insufficiency of specification or performance insufficiency}. Finally, the definition of hazard is adapted from the given in the ISO26262,~\textit{“potential source of harm caused by malfunctioning behaviour of the item”}. SOTIF standard replaces the word "malfunctioning" by "hazardous" and the phrase "of the item" by "at the vehicle level" in comparison with the given by the ISO26262 to adapt the definition to the scope of the standard. For clarification, the insufficiencies of specification are out of the scope of the related project to this work, therefore, a functional insufficiency is considered the same as a functional insufficiency in this work as is shown in the box of the project scope in Figure~\ref{fig:masa_cause_and_effect_model}. The inclusion of a triggering condition could start a reaction in the system that could activate a functional insufficiency and could finally result in a hazardous behaviour. A hazardous behaviour is defined based on the result of the Key Performance Indicators (KPIs) or Safety Performance Indicator (SPIs). These metrics are used to discern if the result of the tests is within a defined tolerable value or, on the other hand, is outside the tolerable windows and is set as a hazardous behaviour. A KPI is a metric that is used for measure a specific parameter of the system. In a similar way, a SPI defines a metric but focused on the safety domain such as the Minimum Safe Distance Violation (MSDV) or the Time-to-Collision (TTC)~\cite{self_driving_car_safety}

	The Operational Design Domain (ODD) is a crucial concept in the scenario validation process. While it is defined in the SOTIF standard, the definition provided in the UL4600 standard~\cite{UL4600} outlines more precisely what an ODD constitutes from our perspective. As per this standard, an ODD refers to \textit{"the set of environments and situations the item is intended to operate within. This includes not only direct environmental conditions and geographic restrictions, but also a characterization of the set of objects, events, and other conditions that will occur within that environment"}. 
	The scenario development for the ODDs utilises the widely accepted methodology of The 6-Layer Scenario Model~\cite{SAEJ3016}\cite{layer_scenario_model}. This model splits the definition of each scenario into six layers, each concentrating on the context of the scenario. The layers and their definitions are: 
	
	\begin{itemize}
		\item \textit{Layer 1 – Road network and traffic guidance objects}: e.g. road markings, and traffic signs and traffic lights.
		\item \textit{Layer 2 – Roadside structures}: e.g. buildings, vegetation, streets lamps, and advertising boards.
		\item \textit{Layer 3 – Temporary modifications of L1 and L2}: e.g. roadwork signs, temporary markings, and covered markings.
		\item \textit{Layer 4 – Dynamic Objects}: e.g. vehicles (moving and non-moving), pedestrians (moving and non-moving), trailers, and animals.
		\item \textit{Layer 5 – Environmental Conditions}: e.g. illumination, precipitation, and road weather.
		\item \textit{Layer 6 – Digital information}: e.g. state of traffic lights, switchable traffics signs, and V2X messages. 
	\end{itemize}

	In this context, attempts have been made to define a taxonomy that can describe most scenarios in the most detailed manner. For example, the one from the British Standards Institution~\cite{BSI} or the taxonomy from the Society of Automotive Engineers (SAE)~\cite{SAEAVSC} are widely recognized. Additionally, Annex B of the SOTIF standard also addresses this matter.

	Once the main concepts have been outlined, the following section describes the integration process into the validation tool. Finally, section~\ref{sec:conclusions} provides a summary of all the work presented in this publication and establishes the direction for future research. 
	
\begin{figure*}
	\centering
	\includegraphics[width=0.8\textwidth]{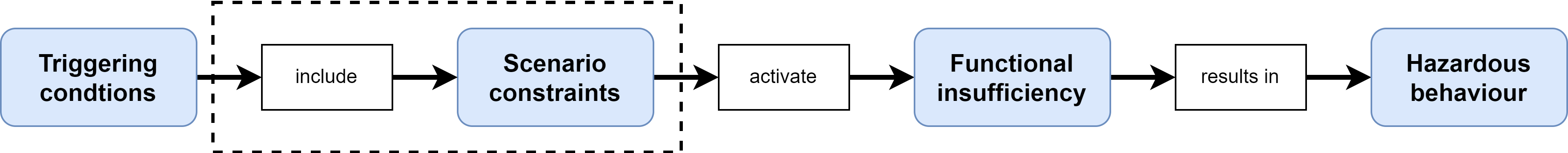}
	\caption{Updated cause-and-effect model including the scenarios constraints}
	\label{fig:sense_plan_act_model_scenario_constrain}
\end{figure*}

\begin{figure}
	\centering
	\includegraphics[width=0.40\textwidth]{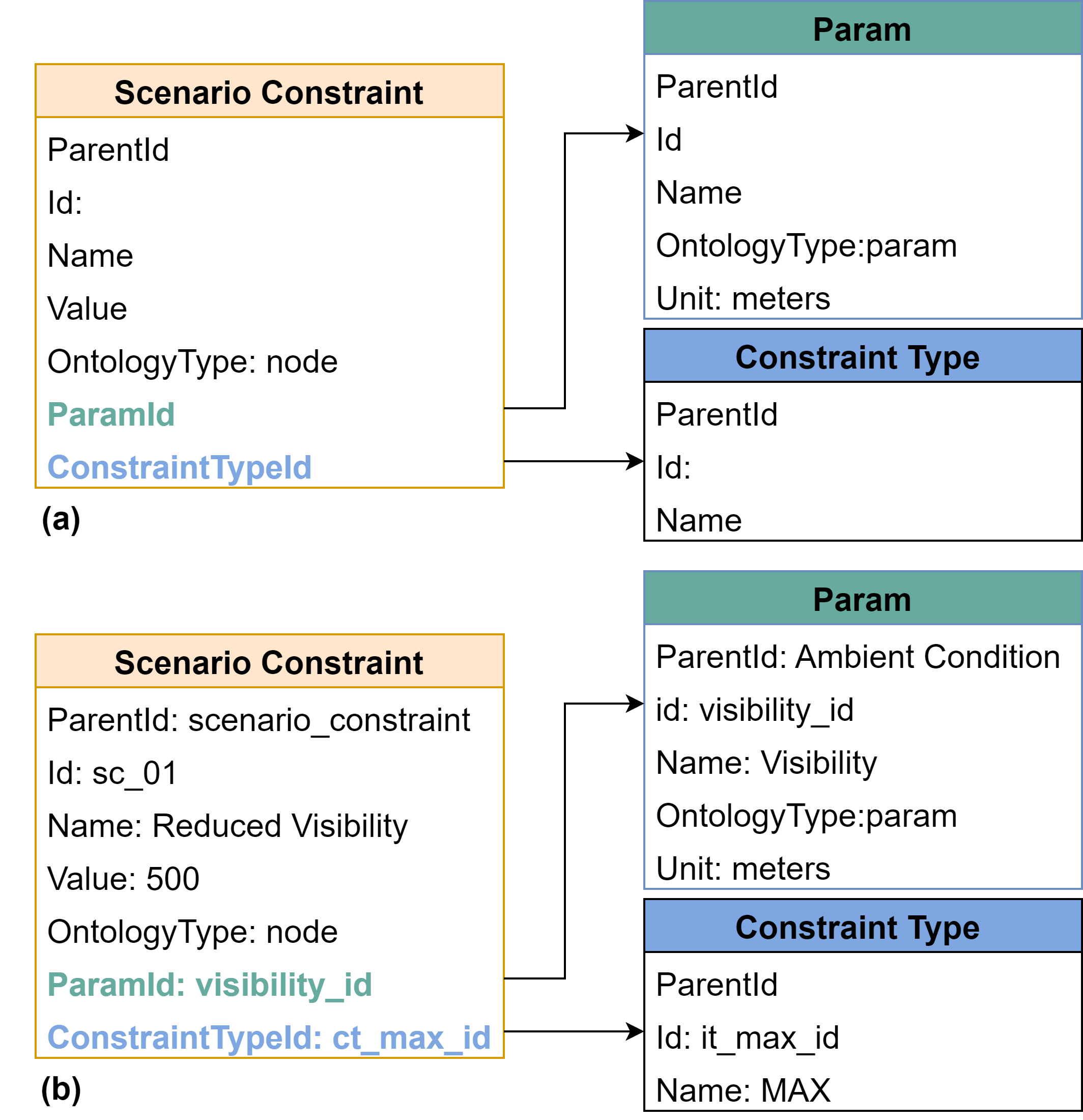}
	\caption{Definition of a scenario constraint}
	\label{fig:scenario_constraint_definition}
\end{figure}

\begin{figure*}
	\centering
	\includegraphics[width=0.85\textwidth]{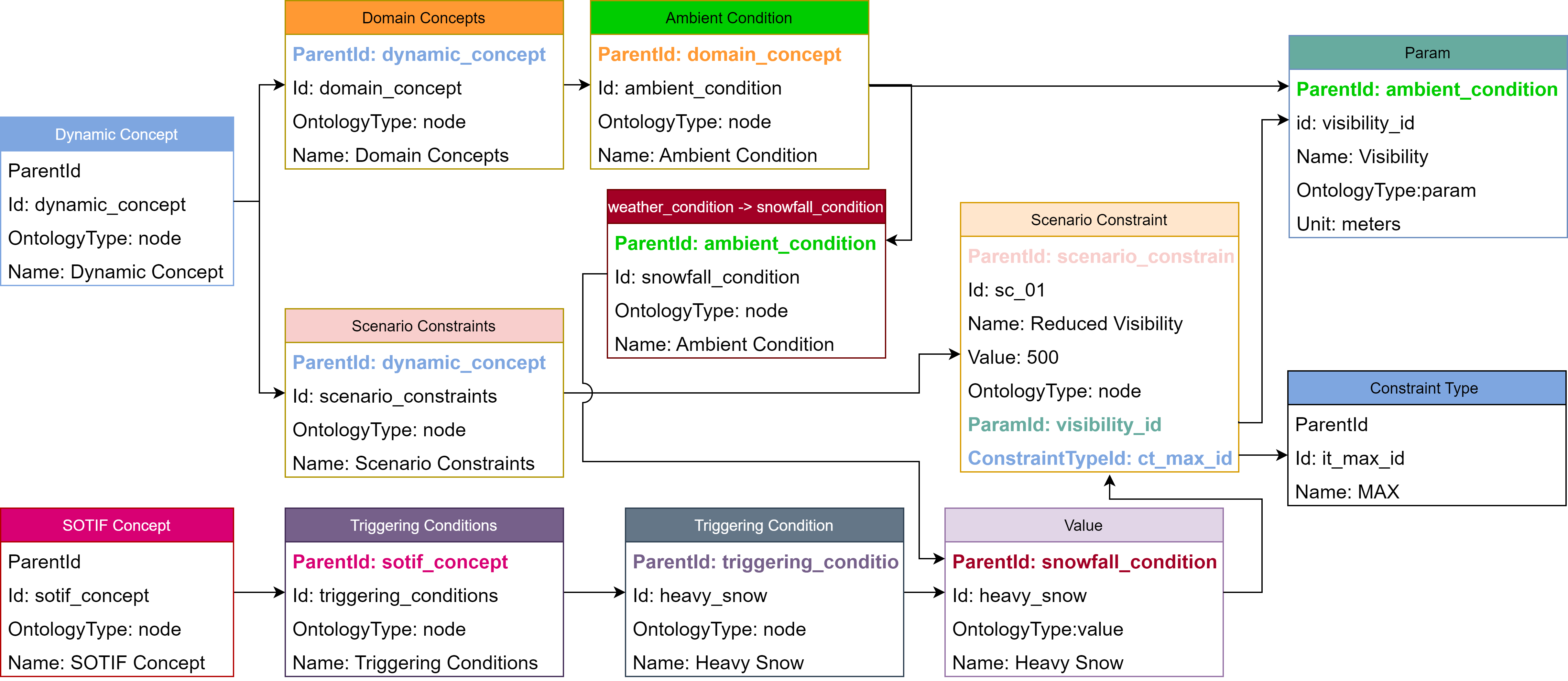}
	\caption{Heavy Snow triggering condition in SCENIUS tool suite}
	\label{fig:heavy_snow_tc_in_scenius}
\end{figure*} 

\begin{figure}
	\centering
	\includegraphics[width=0.4\textwidth]{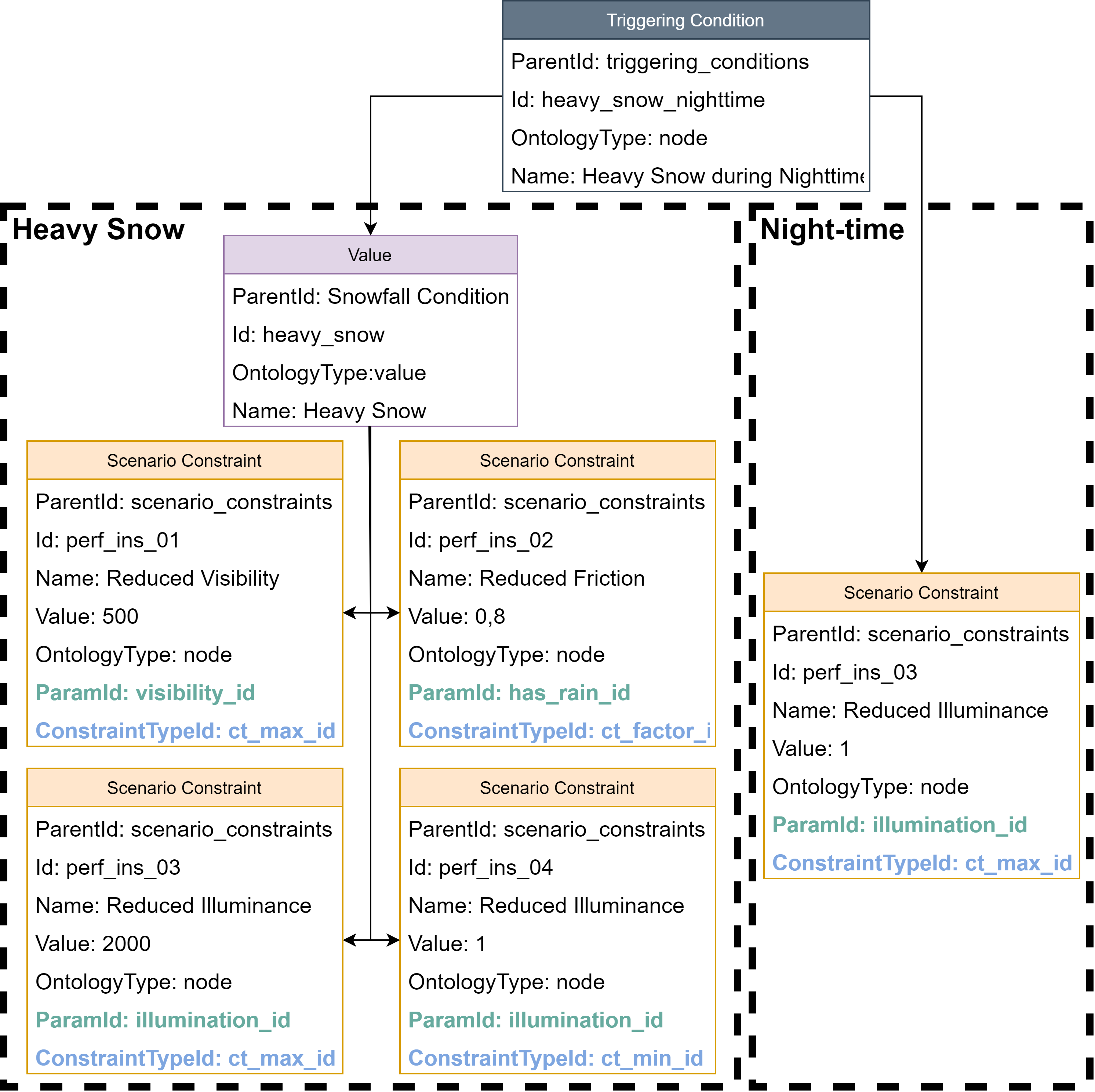}
	\caption{Extended Triggering Condition: "Heavy Snow during Night-time"}
	\label{fig:extended_tc}
\end{figure}

\section{SOTIF Concept Integration Process}
\label{sec:sotif_concept}	
	The integration has been implemented within the AVL SCENIUS\textsuperscript{TM}~\cite{avl_scenius} tool suite, was developed with the scenario-based validation approach in mind. The complete validation process is covered, from scenario design to scenario management, test case generation, test allocation, and result reporting. The suite is based on three main tools modules. First, the scenario designer allows the user to handle all aspects of the scenario including the parametrisation. It fully supports ASAM OpenScenario~\cite{asam_open_scenario} and OpenDrive~\cite{asam_open_drive}. All scenarios are immediately verified for standard conformity as well as by the enhanced data and logic checks. Then, the user could manage all stored scenarios in the scenario data manager. All the elements relevant for the sufficient description of scenarios such as road content, traffic content, and other environmental data are managed and stored in a central database. Finally, the test case generator provides the user the possibility of defining test orders in a simulation or transfer to another different execution environment. The implemented smart testing algorithms enable the automatic reduction of the vast amount of test cases and parameter variations. In addition to the main benefits provided by the tool suite such as time-cost saving, efficiency, fast integration and traceability; the inclusion of the SOTIF concept extends and improves the identification and validation of both unknown and known hazardous scenarios of a ADAS/AD function to obtain a more precise safety argumentation. 	
	
	An ontology is used to describe the scenarios that will be used for testing. ASAM OpenXOntology~\cite{asam_open_ontology} is used as a reference, but modifications are included to better fit the requirements of the tool chain. Internally, the ontology and its relationships are defined by using four kind of entities:

	\begin{itemize}
		\item \textit{Node:} A node is the entity in which the hierarchy of the ontology is built. It can be a child of another node, or a parent for Enums or Params. Examples of nodes are the ambient or weather conditions of the scenario, which are the parents of scenario parameters such as rain or illumination parameter.
		\item \textit{Enum:} Defines a list of values that are related to each other. For example, an enum is the snowfall condition, which is defined by three different levels of severity: heavy snow, light snow, and moderate snow. 
		\item \textit{Value:} Defines an entity that is an abstraction of a phenomenon, but it is not yet modelled. In this case, the phenomenon already exits in the system but has not yet been parametrised. As example, this type of entity are the one previously mentioned: heavy snow, light snow, and moderate snow.
		\item \textit{Param:} Defines an entity that can be quantified. Each one is associated with a unit to be measured. For example, scenario illuminance, which is associated with lux units.
	\end{itemize}

	The integration of the SOTIF concept needs the addition of a new node in the system, which is the parent of all triggering condition defined in the validation tool. Next integration step is to linking the triggering conditions with the existing defined ontology to be able to parametrised each triggering condition and included it into the test scenario. Therefore, an intermediate block has to be added to the cause-and-effect model, which is shown in Figure~\ref{fig:sense_plan_act_model_scenario_constrain}. In this new approach, a scenario constraint block is added, which connects the triggering conditions with the performance insufficiencies. By using this approach, we are able to define each triggering conditions as a combination of one or many scenario constraints. Following the scenario constraint parametrisation shown in Figure~\ref{fig:heavy_snow_tc_in_scenius}(a), a scenario constraint is a node entity that is linked to a \textit{Param} and a new entity called \textit{Constraint Type}, which sets the type of constraint such as a maximum or a minimum value. The \textit{Param} entity is associated to the existing ontology of the system. According to BSI PAS 1883 standard~\cite{BSI}, a heavy snow condition is defined as a visibility limitation up to 500 meters. This scenario constraints is defined in Figure~\ref{fig:heavy_snow_tc_in_scenius}(b), where relationship and value is given in the \textit{Node} entity that is associated to the \textit{Param} from the ontology (visibility) and the type of constraint (MAX). The complete hierarchy and relationship tree of the system for this specific triggering condition is given in Figure~\ref{fig:heavy_snow_tc_in_scenius}.

	The SOTIF concept is designed with scalability in mind due to the test scenario shall increase the complexity of the triggering conditions and their parametrization. An example of this increasing complexity, the triggering condition \textit{Heavy Snow during Night-time} is shown in Figure~\ref{fig:extended_tc}. These types of triggering conditions are treated as the combination of two independent triggering conditions:~\textit{Heavy Snow + Night-time}. In contrast to the previous example, the weather condition \textit{Heavy Snow} is more finely parametrised. In this context, not only is the impact in the visibility is considered, but also the effect on the scenario illumination and the asphalt friction. Following the standard again, the illuminance in a heavy snow scenario can be parametrised from 1 lux to 2000 lux. Additionally, a reduction factor of 0.8 is applied on the asphalt friction in this potential triggering condition. However, this particular triggering condition occurs during night-time, therefore, the illuminance condition due to night-time is also applied (illuminance less than 1 lux according to the standards). Therefore, in this triggering condition parametrization there are two illuminance constraints. In this situation, the most limiting conditions is applied. It means that in this definition, the illuminance parametrised in night-time overrides the value of the illuminance parametrised in the heavy snow condition. 
	
	Finally, when one or more potential triggering conditions are selected in the scenario for testing, the scenario constraints (e.g., limited visibility, reduced friction...) associated to each potential triggering condition are also included in the generated test cases. The resulting metrics of the matrix test cases show the impact of the selected potential triggering conditions on the function, which are compared with the nominal performance of the function (i.e., no potential triggering conditions included) to determine them not longer as potential but triggering conditions for the function, and to identify the thresholds at which they are relevant to impact and effect on the function output.
	
\section{Conclusions and future work}
\label{sec:conclusions}

	In this publication, the integration of a SOTIF concept has been explained, where some adaptations and parametrisation of the scenario constraints have to be done in order to integrate triggering conditions into an existing scenario ontology. As a first step, an extensive list of potential triggering conditions has been investigated based on current state-of-the-art and available standards. They are then parametrised by using an existing system ontology, which is used to model the scenarios and defining the needed entities and relationship to be able to  link the triggering conditions, scenario constraints, and the existing ontology.
		
	As a future task, we will define the triggering conditions that cannot be parametrised using existing standards. Moreover, we are researching a methodology that allows us to capture the majority of potential triggering conditions based on the performance insufficiencies on the perception side. This is due to the infinite number of triggering conditions in the real world, which are not possible to cover manually. 
	
\section{Acknowledgement}
\label{sec:acknowledgements}

The publication was written at Virtual Vehicle Research GmbH in Graz, Austria. The authors would like to acknowledge the financial support within the COMET K2 Competence Centers for Excellent Technologies from the Austrian Federal Ministry for Climate Action (BMK), the Austrian Federal Ministry for Labour and Economy (BMAW), the Province of Styria (Dept. 12) and the Styrian Business Promotion Agency (SFG). The Austrian Research Promotion Agency (FFG) has been authorised for the programme management. They would furthermore like to express their thanks to their supporting industrial project partner, namely AVL List GmbH.

\bibliographystyle{ieeetr}
\bibliography{bibliography}
\balance

\end{document}